\documentstyle[twocolumn,pre,aps]{revtex}
\begin{document}
\draft

\title{
Nonequilibrium relaxation of the two-dimensional Ising model:
Series-expansion and Monte Carlo studies
}

\author{Jian-Sheng Wang$^\dag$ and Chee Kwan Gan$^{\dag\,\ddag}$}

\address{
$^\dag$Department of Computational Science, \\
National University of Singapore, Singapore 119260,\\
Republic of Singapore. \\
$^\ddag$Cavendish Laboratory, University of Cambridge, Madingley Road, \\
Cambridge CB3 0HE, United Kingdom.
}

\date{17 December 1997}

\maketitle

\begin{abstract}
We study the critical relaxation of the two-dimensional Ising
model from a fully ordered configuration by series expansion in
time $t$ and by Monte Carlo simulation.  Both the magnetization
($m$) and energy series are obtained up to 12-th order.  An
accurate estimate from series analysis for the dynamical critical exponent
$z$ is difficult but compatible with 2.2.  We also use Monte
Carlo simulation to determine an effective exponent,
$z_{\text{eff}}(t) = - {1 \over 8} d \ln t /d \ln m $, directly
from a ratio of three-spin correlation to $m$.  Extrapolation to
$t\to\infty$ leads to an estimate $z = 2.169 \pm 0.003$.
\end{abstract}


\narrowtext
\section{Introduction}

The pure relaxational dynamics of the kinetic Ising model with no
conserved fields, which is designated as model A in the
Hohenberg-Halperin review
\cite{Hohenberg-Halperin-77},
has been studied extensively by various approaches.  Unlike some
of the other models in which the dynamical critical exponent $z$
can be related to the static exponents, it seems that $z$ of
model A is independent of the static exponents (however, see
Ref.~\cite{Alexandrowicz-90}).  In the past twenty years, the
numerical estimates for the dynamical critical exponent $z$
scattered a lot, but recent studies seem to indicate a
convergence of estimated values.  Our studies contribute further to
this trend.

We review briefly some of the previous work on the computation of
the dynamical critical exponents, concentrating mostly on the
two-dimensional Ising model.  The conventional theory predicts $z
= 2 -\eta$ \cite{vanHove-54}, where $\eta$ is the critical
exponent in the two-point correlation function, $G(r)
\propto r^{-d+2-\eta}$.  For the two-dimensional Ising model, this gives
$z=1.75$.  It is known that this is only a lower bound
\cite{Abe-69}.  It is very interesting to note that series
expansions\cite{Yahata-Suzuki-69,Yahata-71,Racz-Collins-76,%
Rogiers-Indekeu-90,Wang-93,Dammann-Reger-93} gave one of the
earliest quantitative estimates of $z$.  Dammann and
Reger\cite{Dammann-Reger-93} have the longest high-temperature
series (20 terms) for the relaxation times so far, obtaining
$z=2.183 \pm 0.005$.  However, re-analysis of the series by
Adler\cite{Adler-96} gives $z=2.165 \pm 0.015$.  There are two
types of field-theoretic renormalization group analysis: the
$\epsilon$-expansion near dimension $d=4$
\cite{Halperin-72,DeDominicis-75} and an interface model
near $d=1$ \cite{Bausch-81}.
It is not clear how reliable when it is interpolated to $d=2$.
Real-space renormalization group of various schemes has been
proposed in the early eighties
\cite{Achiam-80,Mazenko-Valls-81,Takano-Suzuki-82,Haake-Lewenstein-Wilkens-84},
but it appears that there are controversies as whether some of
the schemes are well-defined.  The results are not of high
accuracy compared to other methods.  Dynamic Monte Carlo
renormalization group
\cite{Tobochnik-81,Kalle-84,Williams-85,Lacasse-93}  
is a generalization of the equilibrium Monte Carlo
renormalization group method
\cite{MCRG}.  The latest work
\cite{Lacasse-93} gives $z=2.13\pm 0.01$ in two dimensions. 
Equilibrium Monte Carlo method is one of the standard methods to
estimate $z$ \cite{Stoll-Binder-Schneider-73,%
Tang-Wansleben-87,Landau-94,Pearson-85,Heuer-92}.  However, long
simulations ($t \gg L^z$) are needed for sufficient statistical
accuracy of the time-displaced correlation functions.  The
analysis is quite difficult due to unknown nature of the
correlation functions.  Nonequilibrium relaxation
\cite{Kikuchi-Okabe-86,Bonfim-87,%
Mori-Tsuda-88,Stauffer-92,Muenkel-93,Ito-93}, starting from a
completely ordered state at $T_c$, has nice features.  The
analysis of data is more or less straightforward.  The lattice
can be made very large, so that finite-size effect can be ignored
(for $t \ll L^z$).  The catch here is that correction to scaling
due to finite $t$ is large.  Recently, the idea of damage
spreading
\cite{Stauffer-93,Matz-94,Grassberger-95,Gropengiesser-95,%
Wang-Suzuki-Hu-95} has also been employed.  Methods based on
statistical errors in equilibrium Monte Carlo simulation
\cite{Kikuchi-Ito-Okabe-94}, finite-size scaling of
nonequilibrium relaxation
\cite{Li-Schuelke-Zhang-95,Soares-97},  
and finite-size scaling of the eigenvalues of the stochastic
matrix
\cite{Pandit-Forgacs-Rujan-81,Nightingale-Bloete-96} are used to 
compute the exponent.  A recent calculation with a
variance-reducing Monte Carlo algorithm for the leading
eigenvalues gives prediction \cite{Nightingale-Bloete-96}
$z=2.1665 \pm 0.0012$.  This appears to be the most precise value
reported in the literature.

The high-temperature series expansions for the relaxation times
are often used in the study of Ising dynamics.  In this paper, we
present a new series which directly corresponds to the
magnetization (or energy) relaxation at the critical temperature.
Our series expansion method appears to be the only work which
uses time $t$ as an expansion parameter.  The generation of these
series is discussed in Sections~\ref{sec:expansion}
and~\ref{sec:comp}.  Dynamical scaling mentioned in
Section~\ref{sec:scaling} forms the basis of the analysis, and
the results are analyzed in Section~\ref{sec:analysis}.  We feel
that the series are still too short to capture the dynamics at
the scaling regime.  We also report results of an extensive Monte
Carlo simulation for the magnetization relaxation.  We find that
it is advantageous to compute an effective dynamical critical
exponent directly with the help of the governing master equation
(or the rate equation).  The simulation and analysis of Monte
Carlo data are presented in Section~\ref{sec:simulation}.  We
summarize and conclude in Section~\ref{sec:conclusion}.

\section{Series expansion method}
\label{sec:expansion}

In this section, we introduce the relevant notations, and outline
our method of series expansion in time variable $t$.  The
formulation of single-spin dynamics has already been worked out
by Glauber
\cite{Glauber-63}, and by Yahata and Suzuki\cite{Yahata-Suzuki-69} long time
ago.  To our knowledge, all the previous series studies for Ising
dynamics
\cite{Yahata-Suzuki-69,Yahata-71,Racz-Collins-76,%
Rogiers-Indekeu-90,Wang-93,Dammann-Reger-93} are based on
high-temperature expansions of some correlation times.  
As we will see, expansion in $t$ is
simple in structure, and it offers at least a useful alternative for
the study of Ising relaxation dynamics.

We consider the standard Ising model on a square lattice
\cite{Onsager-44} with the energy of a configuration $\sigma$
given by
\begin{equation}
  E(\sigma) = - J \sum_{\langle i,j\rangle} \sigma_i \sigma_j,
\end{equation}
where the spin variables $\sigma_i$ take $\pm 1$, $J$ is the
coupling constant, and the summation runs over all nearest
neighbor pairs.  The thermal equilibrium value of an observable
$f(\sigma)$ at temperature $T$ is computed according to the
Boltzmann distribution,
\begin{equation}
 \langle f \rangle = { \sum_{\sigma} f(\sigma) 
         \exp\bigl(- E(\sigma)/k_B T\bigr) \over
           \sum{_\sigma}  \exp\bigl(- E(\sigma)/k_B T\bigr) } 
                   = \sum_{\sigma} f(\sigma) P_{\text{eq}}(\sigma).
\end{equation}
The equilibrium statistical-mechanical model defined above has no
intrinsic dynamics.  A local stochastic dynamics can be given and
realized in Monte Carlo simulations
\cite{MullerKrumbhaar-73}.
The dynamics is far from unique; in particular, cluster dynamics
\cite{Swendsen-Wang-87} differs vastly from the local ones.

A sequence of Monte Carlo updates can be viewed as a discrete
Markov process.  The evolution of the probability distribution is
given by
\begin{equation}
  P(\sigma, k+1) = \sum_{\sigma'}P(\sigma', k) W(\sigma'| \sigma),
\end{equation}
where $W$ is a transition matrix satisfying the stationary
condition with respect to the equilibrium distribution, i.e.,
$P_{\text{eq}} = P_{\text{eq}} W$.  A continuous time description
is more convenient for analytic treatment.  This can be obtained
by fixing $t = k/N$, and letting $\delta t = 1/N \to 0$, where
$N=L^2$ is the number of spins in the system.  The resulting
differential equation is given by
\begin{equation}
  {\partial P(\sigma, t) \over \partial t} = \Gamma P(\sigma, t), 
\label{mastereq}
\end{equation}
where $\Gamma$ is a linear operator acting on the vector
$P(\sigma,t)$, which can be viewed as a vector of dimension
$2^N$, indexed by $\sigma$.  If we use the single-spin-flip
Glauber dynamics \cite{Glauber-63}, we can write
\begin{equation}
  \Gamma = -\sum_{j=1}^N w_j(\sigma_j) + \sum_{j=1}^N
w_j(-\sigma_j)F_j,
\end{equation}
where 
\begin{equation}
  w_j(\sigma_j) = { 1\over 2}\biggl[ 1 - \sigma_j \tanh\Bigl( K\!\! 
                   \sum_{\text{nn of j}}\!\!\sigma_k\Bigr) \biggr],
\; K = {J \over k_BT},
\label{glauber}
\end{equation}
and $F_j$ is a flip operator such that
\begin{equation}
F_j P(\ldots,\sigma_j, \ldots) = P(\ldots,-\sigma_j, \ldots).
\end{equation}
The flip rate $w_j(\sigma_j)$ for site $j$ depends on the spin
value at the site $j$ as well as the values of its nearest
neighbor spins $\sigma_k$.

The full probability distribution clearly contains all the
dynamic properties of the system.  Unfortunately its high
dimensionality is difficult to handle.  It can be shown from the
master equation, Eq.~(\ref{mastereq}), that any function of the
state $\sigma$ (without explicit $t$ dependence) obeys the
equation
\begin{equation}
  {d\langle f\rangle_t \over dt} = - \langle {\cal L} f \rangle_t,
\label{rateeq}
\end{equation}
where 
\begin{equation}
{\cal L} = \sum_{j=1}^N w_j(\sigma_j) (1 - F_j),
\end{equation}
and the average of $f$ at time $t$ is defined by
\begin{equation}
  \langle f \rangle_t = \sum_{\sigma} f(\sigma) P(\sigma, t).
\end{equation}
Note that the time dependence of $\langle f\rangle$ is only due to $P(\sigma,
t)$.  For the series expansion of this work, it is sufficient to
look at a special class of functions of the form $\sigma^A =
\prod_{j \in A} \sigma_j$, where $A$ is a set of sites.  In such
a case we have
\begin{equation}
  {d\langle \sigma^A \rangle_t \over dt} = - 2 
 \sum_{j \in A} \bigl\langle w_j(\sigma_j) \sigma^A \bigr\rangle_t.
\label{rateeq2}
\end{equation}
With this set of equations, we can compute the $n$-th derivative
of the average magnetization $\langle\sigma_0\rangle_t$.  A
formal solution to Eq.~(\ref{rateeq}) is
\begin{equation}
  \langle \sigma^A \rangle_t = 
\left\langle e^{ - {\cal L} t} \sigma^A \right\rangle_0 
= \sum_{n=0}^\infty \left\langle { (-{\cal L} t)^n \over n!} \sigma^A \right\rangle_0.
\end{equation}
This equation or equivalently the rate equation,
Eq.~(\ref{rateeq2}), forms the basis of our series expansion in
time $t$.

A few words on high-temperature expansions are in order here.
They are typically done by integrating out the time
dependence---the nonlinear relaxation time can be defined as
\begin{equation}
  \tau_{\text{nl}}^A = \int_0^\infty \langle \sigma^A \rangle_t dt =
\left\langle\!\int_0^\infty dt\, e^{ - {\cal L} t} \sigma^A \right\rangle_0 
 = \langle {\cal L}^{-1} \sigma^A \rangle_0.
\end{equation}
The equilibrium correlation time (linear relaxation time) can be
expressed as
\begin{eqnarray}
\tau = && \int_0^\infty { \langle m(0)m(t)\rangle_{\text{eq}} \over 
                                    \langle m(0)^2\rangle_{\text{eq}}}
dt =\sum_{j}\int_0^\infty \langle \sigma_0 e^{-{\cal L}t} \sigma_j \rangle_{\text{eq}}
/\chi \nonumber \\
 = && {1\over \chi} \sum_j 
  \langle \sigma_0 {\cal L}^{-1} \sigma_j \rangle_{\text{eq}}.
\end{eqnarray}
where $\chi = N \langle m^2 \rangle_{\text{eq}}$ is the reduced
static susceptibility.  The average is with respect to the
equilibrium distribution, $P_{\text{eq}}(\sigma)$.  A suitable
expansion in small parameter $J/k_BT$ can be made by writing
${\cal L} = {\cal L}_0 + \Delta {\cal L}$.

It is clear that we can also perform the Kawasaki dynamics with a
corresponding rate.  Of course, since the magnetization is
conserved, only energy and higher order correlations can relax.

A very convenient form for the Glauber transition rate,
Eq.~(\ref{glauber}), on a two-dimensional square lattice is
\begin{eqnarray}
  w_0(\sigma_0) = && {1\over2} \Big[ 1 + x \sigma_0 (\sigma_1 \!+\! 
\sigma_2 \!+\! \sigma_3 \!+\! \sigma_4) \nonumber\\
 && + y \sigma_0 (     
\sigma_1 \sigma_2 \sigma_3 \!+\! \sigma_2 \sigma_3 \sigma_4 \!+\! 
\sigma_3 \sigma_4 \sigma_1 \!+\! \sigma_4 \sigma_1 \sigma_2 ) \Big],
\label{ratesqlat}
\end{eqnarray}
\begin{equation}
   x = - {1\over 4} \tanh 2K - {1\over 8} \tanh 4K,
\end{equation}
\begin{equation}
   y = + {1\over 4} \tanh 2K - {1\over 8} \tanh 4K,
\end{equation}
where the site 0 is the center site, and sites 1, 2, 3, and 4 are
the nearest neighbors of the center site.  At the critical
temperature, $\tanh K_c = \sqrt{2} -1 $, we have $x = - 5
\sqrt{2}/24$ and $y = \sqrt{2}/24$.

\section{Computer implementation and results}       
\label{sec:comp}
A series expansion in $t$ amounts to finding the derivatives
evaluated at $t=0$:
\begin{equation}
 \langle \sigma^A\rangle_t = \sum_{n=0}^\infty { t^n\over n!} 
{ d^n\langle \sigma^A\rangle \over dt^n }\Big|_{t=0}.
\end{equation}
The derivatives are computed using Eq.~(\ref{rateeq2})
recursively.  A general function is coded in C programming
language to find the right-hand side of Eq.~(\ref{rateeq2}) when
the configuration $\sigma^A$, or the set $A$, is given.  The set
$A$ is represented as a list of coordinates constructed in an
ordered manner. By specializing the flip rate as given by
Eq.~(\ref{ratesqlat}), and considering each site in $A$ in turn,
the configurations on the right-hand side of the rate equation
are generated in three ways: (1) the same configuration as $A$,
which contributes a factor (coefficient of a term) of $-1$; (2) a
set of configurations generated by introducing a pair of nearest
neighbor sites in four possible directions, with one of the sites
being the site in $A$ under consideration, and making use of the
fact $\sigma_i^2=1$. We notice that the site in $A$ under
consideration always gets annihilated.  Each resulting
configuration contributes a factor of $-x$; and (3) same as in
(2) but two more sites which are also the nearest neighbors of
the site in $A$ under consideration are introduced.  This two
extra sites form a line perpendicular to the line joined by the
first pair of neighbor sites in (2).  Each of this configuration
has a factor of $-y$.  It is instructive to write down the first rate
equation, taking into account of the lattice symmetry (e.g.
$\langle \sigma_i \rangle = \langle \sigma_0
\rangle$, for all $i$):
\begin{equation}
{ d \langle \sigma_0 \rangle \over dt } = -(1+4x) \langle \sigma_0 \rangle 
 - 4 y \langle \sigma_1\sigma_2\sigma_3 \rangle. \label{first-rate}
\end{equation}

The core of the computer implementation for series expansion
\cite{Gan-Wang}              
is a symbolic representation of the rate equations.  Each rate
equation is represented by a node together with a list of
pointers to other nodes.  Each node represents a function
$\langle \sigma^A\rangle$, and is characterized by the set of
spins $A$.  The node contains pointers to the derivatives of this
node obtained so far, and pointers to the ``children'' of this
node and their associated coefficients, which form a symbolic
representation of the rate equations.  The derivatives are
represented as polynomials in $y$.  Since each node is linked to
other nodes, the computation of the $n$-th derivative can be
thought of as expanding a tree (with arbitrary number of
branches) of depth $n$.

The traversal or expansion of the tree can be done in a
depth-first fashion or a breadth-first fashion.  Each has a
different computational complexity.  A simple depth-first
traversal requires only a small amount of memory of order $n$.
However, the time complexity is at least exponential, $b^n$, with
a large base $b$.  A breadth-first algorithm consumes memory
exponentially, even after the number of the rate equations has
been reduced by taking the symmetry of the problem into account.
The idea of dynamic programming can be incorporated in the
breadth-first expansion where the intermediate results are stored
and referred.  To achieve the best performance, a hybrid of
strategies is used to reduce the computational complexity:

\begin{itemize}
\item Each configuration (pattern) is transformed into  
its canonical representation, since all configurations related by
lattice symmetry are considered as the same configuration.
\item We use breadth-first expansion to avoid repeated computations 
involving the same configuration. If a configuration has already
appeared in earlier expansion, a pointer reference is made to the
old configuration.  Each configuration is stored in memory only
once.  However, storing of all the distinct configurations leads
to a very fast growth in memory consumption.
\item The last few generations in the tree expansion
use a simple depth-first traversal to curb the problem of memory
explosion.
\item Parallel computation proves to be useful.
The longest series is obtained by a cluster of 16 Pentium Pro PCs
with high speed network connection (known as Beowulf).
\end{itemize}

The program is controlled by two parameters $D$ and $C$.  $D$ is
the depth of breadth-first expansion of the tree.  When depth $D$
is reached, we no longer want to continue the normal expansion in
order to conserve memory.  Instead, we consider each leave node
afresh as the root of a new tree.  The derivatives up to
$(n-D)$th order are computed for this leave node.  The expansion
of the leave nodes are done in serial, so that the memory
resource can be reused.  The parameter $C$ controls the number of
last $C$ generations which should be computed with a simple
depth-first expansion algorithm.  It is a simple recursive
counting algorithm, which uses very little memory, and can run
fast if the depth $C$ is not very large.  In this algorithm the
lattice symmetry is not treated.  The best choice of parameters
is $D=6$ and $C=2$ on a DEC AlphaStation 250/266.  The computer
time and memory usage are presented in
Table~\ref{tab:complexity}.  As we can see from the table, each
new order requires more than a factor of ten CPU time and about
the same factor for memory if memory is not reused.  This is the
case until the order $D+C+1$, where no fresh leave-node expansion
is made.  There is a big jump (a factor of 60) in CPU time from
9-th order to 10-th order, but with a much smaller increase in
memory usage.  This is due to the change of expansion strategy.
Finally the longest 12-th order series is obtained by parallel
computation on a 16-node Pentium Pro 200 MHz cluster in 12 days.
The number of distinct nodes generated to order $n$ is roughly
${1\over 100} 11^n$.  To 12-th order, we have examined about
$10^{10}$ distinct nodes.  The series data are listed in
Table~\ref{tab:series}.

\section{Dynamical scaling}
\label{sec:scaling}
The traditional method of determining the dynamical critical
exponent $z$ is to consider the time-displaced equilibrium
correlation functions.  However, one can alternatively look at
the relaxation towards thermal equilibration.  The basic
assumption is the algebraic decay of the magnetization at $T_c$,
\begin{equation}
   \langle\sigma_0\rangle = m \approx c\, t^{-\beta/ \nu z}, \qquad t\to\infty.
\label{scaling}
\end{equation}
This scaling law can be obtained intuitively as follows.  Since
the relaxation time and the correlation length are related
through $\tau \propto \xi^z$ by definition, after time $t$, the
equilibrated region is of size $t^{1/z}$.  Each of such
region is independent of the others, so the system behaves as a
finite system of linear length $\xi \propto t^{1/z}$.  According
to finite-size scaling \cite{Privman-90}, the magnetization is of order
$\xi^{-\beta/\nu}$ on a finite system of length $\xi$.  Each
region should have the same sign for the magnetization since we
started the system with all spins pointing in the same direction.
The total magnetization is equal to that of a correlated region,
giving $m \propto t^{-\beta/\nu z}$.

The same relation can be derived from a more general scaling
assumption
\cite{Suzuki-77},
\begin{equation}
  m(t, \epsilon) \approx \epsilon^\beta \phi(t \epsilon^{\nu z}), 
    \qquad \epsilon = { T - T_c \over T_c}.
\end{equation}
By requiring that $m(t, \epsilon)$ is still finite as the scaling
argument $t \epsilon^{\nu z} \to 0$ and $\epsilon \to 0$ with
fixed $t$, we get Eq.~(\ref{scaling}).

Equation (\ref{scaling}) is only true asymptotically for large
$t$.  It seems that there is no theory concerning leading
correction to the scaling. As a working hypothesis, we assume
that
\begin{equation}
   m \approx c\, t^{-\beta/ \nu z} ( 1 + b\, t^{-\Delta}).
\end{equation}
The Monte Carlo simulation results as well as current series
analysis seem to support this with $\Delta $ near 1.   Other
possibility might be $z=2$ with logarithmic correction \cite{Domany-84}.

\section{Analysis of series}
\label{sec:analysis}
A general method for extending the range of convergence of a
series is the Pad\'e analysis
\cite{Baker-Graves-Morris-81,Guttmann-89} where a series
is approximated by a ratio of two polynomials.  We first look at
the poles and zeros of the Pad\'e approximants in variable
$s=t/(t+1)$ for $m$.  Since $t$ varies in the range of
$[0,\infty)$, it is easier to look at $s$, which maps the
interval $[0,\infty)$ to $[0,1)$.  There are clusters of zeros
and poles in the $s$-interval $(1,2)$ which corresponds to
negative $t$.  But interval $[0,1)$ is clear of singularities,
which gives us hope for analytic continuation to the whole
interval $[0,1)$.  If we assume the asymptotic behavior $m
\propto t^{-a}$, then $d \ln m/dt = - a/t \approx -a (1-s)$ for
large $t$ or $s\to 1$.  This means that the Pad\'e approximant
should give a zero around $s=1$.  We do observe zeros
near 1. But it is typically a pair of zeros off the real axis
together with a pole at the real axis near 1, or sometimes, only
a pair of real zeros.  These complications make a quantitative
analysis difficult.

Since we know the exact singular point (corresponding to
$t=\infty$), we use the biased estimates by considering the
function
\begin{equation}
  F(t) =  { d \ln m \over d\ln t} \approx -{\beta \over \nu z}.   
\label{Ffunc}
\end{equation}
An effective exponent $z_{\text{eff}}(t)$ is defined by $z_{\text{eff}}(t) =
- \beta/(\nu F(t)) = -1/(8F(t))$.

Again we prefer to use the variable $s$ to bring the infinity to
a finite value 1.  Due to an invariance theorem
\cite{Baker-Graves-Morris-81},
the diagonal Pad\'e approximants in $s$ and $t$ are equal
exactly.  For off-diagonal Pad\'e approximants, $s$ is more
useful since the approximants do not diverge to infinity.

We use methods similar to that of Dickman {\it et al.}
\cite{Dickman-91} and Adler \cite{Adler-96}.  The general idea is
to transform the function $m(t)$ into other functions which one
hopes to be better behaved than the original function.  In
particular, we require that as $t\to
\infty$, the function approaches a constant related to the
exponent $z$.  The first transformation is the
Eq.~(\ref{Ffunc}).  A second family of transformations is
\begin{equation}
  G_p(t) = { d \ln \int_0^t m(t')^p dt' \over d\ln t}  \approx 1 - {p\over 8z},
\label{Gfunc}
\end{equation}
where $p$ is a real positive number.
One can show that the two functions are related by 
\begin{equation}
  F(t) = { 1\over p} \left( G_p(t) - 1 + {d\ln G_p(t)\over d \ln t}\right).
\label{FGrelation}
\end{equation}
The last transform is 
\begin{equation}
  H(t) = F(t) + {1\over \Delta } t { dF \over dt},
\end{equation}
where $\Delta$ is an adjustable parameter, and $F$ can also be
replaced by $G_p$.  If the leading correction to the constant
part is of the form $t^{-\Delta}$, the transformation will
eliminate this correction term.

The transformation of the independent variable $t$ to other
variable is important
to improve the convergence of the Pad\'e approximants.  We
found that it is useful to consider a generalization of the Euler
transform,
\begin{equation}
  u = 1 - { 1 \over (1+t)^\Delta }.
\label{ind-variable}
\end{equation} 
The parameter $\Delta$ is adjusted in such a way to get best
convergence among the approximants.  Since for $t\to \infty$ or
$u\to1$, a Pad\'e approximant near $u=1$ is an analytic function
in $u$, which implies that the leading correlation is of the form
$t^{-\Delta}$.  Note that $\Delta = 1$ corresponds to the Euler
transformation ($u=s$ when $\Delta = 1$).

One of the fundamental difficulties of the transformation method
is that one does not know a priori that a certain transformation
is better than others.  Worst still, we can easily get misleading
apparent convergence among different approximants.  Thus, we need
to be very careful in interpreting our data.  Specifically, we
found that Eq.~(\ref{Ffunc}) gives less satisfactory result than
that of Eq.~(\ref{Gfunc}), where the independent variable $t$ is
transformed into $u$ according to Eq.~(\ref{ind-variable}).
Figure~\ref{fig:Gb} is a plot of all the Pad\'e approximants of
order $[N,D]$, with $N \ge 4$, $D \ge 4$, and $N + D \le 12$, as
a function of the parameter $\Delta$, for $G_1(t=\infty)$.  Good
convergence is obtained at $\Delta = 1.217$ with $z \approx
2.170$.  The estimates $z$ vary only slightly with $p$, at about
0.005 as $p$ varies from 0.5 to 2.  Using $F(t)$ of
Eq.~(\ref{Ffunc}), the optimal value is $\Delta = 1.4$ with $z
\approx 2.26$.  Using the function $H$ does not seem to
improve the convergence.  Even though the value 2.170 seems
to be a very good result, we are unsure of its significance
since there are large deviations of the Pad\'e approximation 
to the function $F(t)$ for $1/t < 0.2$  from
the Monte Carlo result of Fig.~\ref{fig:zeff}.

An objective error estimate is difficult to give.  Estimates from
the standard deviation of the approximants tend to give a very
small error but incompatible among different methods of analysis.
Different Pad\'e approximants are definitely not independent; we
found that $[N,D]$ Pad\'e is almost equal to $[D,N]$ Pad\'e to a
high precision.  A conservative error we quote from the series
analysis is 0.1.

Analysis of the energy series is carried out similarly with $m$
replaced by $\langle \sigma_0 \sigma_1\rangle -
\sqrt{2}/2$, where the constant $\sqrt{2}/2$ is the
equilibrium value.   The large $t$ asymptotic behavior is $t^{-1/z}$ 
\cite{Stauffer-95}.
Both $F$ and $G$ functions give comparable results, better
convergence is obtained for $\Delta > 1$.  The value for $z$ is
about 2.2, but good crossing of the approximants are not
observed.  We feel better analysis method or longer series is
needed.

\section{Monte Carlo simulation}
\label{sec:simulation}
Our motivation for a Monte Carlo calculation was to check the
series result.  It turns out that the data are sufficiently
accurate to be discussed in their own right.  Such an improved
accuracy is achieved by using Eq.~(\ref{first-rate}), which
permits a direct evaluation of the effective exponent
$z_{\text{eff}}(t)$.

We compute the magnetization $m = \langle \sigma_0 \rangle$,
energy per bond $\langle \sigma_0 \sigma_1\rangle$, and the
three-spin correlation $m_3 = \langle \sigma_1\sigma_2\sigma_3
\rangle$ where the three spins are the nearest neighbors of a 
center site having one of the neighbor missing in the product.
With these quantities, the logarithmic derivative,
Eq.~(\ref{Ffunc}), can be computed exactly without resorting to
finite differences.  From Eq.~(\ref{first-rate}) we can write (at
$T=T_c$)
\begin{equation}
  F(t) = { d \ln m \over d \ln t} = - t\left( 1 + {\sqrt{2}\over 6}
\Bigl( {m_3 \over m} - 5\Bigr) \right) = - { 1 \over 8 z_{\text{eff}}(t)}.
\label{ratio}
\end{equation}
The above equation also defines the effective exponent
$z_{\text{eff}}(t)$ which should approach the true exponent $z$ as
$t\to\infty$.

The estimates for the effective exponent based on the ratio of
one spin to three-spin correlation, Eq.~(\ref{ratio}), have
smaller statistical errors in comparison to a finite
difference scheme based on $m(t)$ and $m(t+1)$. Error
propagation analysis shows that the latter has an error 5 times
larger.  Both methods suffer from the same problem that error
$\delta z \propto t$.  Thus, working with very large $t$ does not
necessarily lead to any advantage.

In order to use Eq.~(\ref{ratio}), we need exactly the same flip rate
as in the analytic calculations, namely the Glauber rate,
Eq.~(\ref{glauber}).  The continuous time dynamics corresponds to
a random selection of a site in each step.  Sequential or
checker-board updating cannot directly be compared with the
analytic results.  However, it is believed that the dynamical
critical exponent $z$ does not depend on the details of the
dynamics.

We note that a Monte Carlo simulation is precisely described by a
discrete Markov process while the series expansion is based on
the continuous master equation.  However, the approach to the
continuous limit should be very fast since it is controlled by
the system size---the discreteness in time is $1/L^2$.  We have
used a system of $10^4 \times 10^4$, which is sufficiently large.
Apart from the above consideration, we also checked finite-size
effect.  Clearly, as $t > L^z$, finite-size effect begins to show
up.  We start the system with all spins up, $m(0) = 1$, and
follow the system to $ t = 99$.  For $t<100$, we did not find any
systematic finite-size effect for $L \geq 10^3$.  So the
finite-size effect at $L=10^4$ and $t<100$ can be safely ignored.

Figure~\ref{fig:zeff} shows the Monte Carlo result for the
effective exponent as a function of $1/t$.  The quantities $m$,
$m_3$, and $\langle \sigma_0\sigma_1\rangle$ are averaged over
1868 runs, each with a system of $10^8$ spins.  The total amount
of spin updating is comparable to the longest runs reported in
the literature.  Based on a least-squares fit from $t=30$ to 99,
we obtain
\begin{equation}
  z = 2.169 \pm 0.003.
\end{equation}
The error is obtained from the standard deviation of few groups of
independent runs.  An error estimate based on the residues in the
linear least-squares fit is only half of the above value, which
is understandable since the points in Fig.~\ref{fig:zeff} are not
statistically independent.

In Fig.~\ref{fig:zeff}, we also plot a series result for the
$F(t)$, obtained from the [6,6] Pad\'e of $G_1(u)$ and
Eq.~(\ref{FGrelation}).  Substantial deviations are observed
for  $ 1/t < 0.2$, even though in the $1/t\to0$ limit,
both results  are
almost the same.  This casts some doubts on the reliability of the
series analysis.  We note that the $t\to\infty$ limit of the 
function $F(t)$ is invariant
against any transformation in $t$ which maps $t=\infty$ to $\infty$.
Thus, the discrepancy might be eliminated by a suitable transformation
in the Pad\'e analysis.

\section{Conclusion}
\label{sec:conclusion}

We have computed series for the relaxation of magnetization and
energy at the critical point. The same method can be used to
obtain series at other temperatures or for other correlation
functions. The analyses of the series are non-trivial.  We may
need much more terms before we can obtain result with accuracy
comparable to the high-temperature series.  We have also studied
the relaxation process with Monte Carlo simulation.  The ratio of
three-spin to magnetization is used to give a numerical estimate of
the logarithmic derivatives directly.
This method gives a more accurate estimate for the dynamical
critical exponent.

\section*{Acknowledgment}
This work was supported in part by an Academic Research Grant
No.~RP950601.

\newpage
\bibliographystyle{plain}

\begin{thebibliography}{1}
\bibitem{Hohenberg-Halperin-77}
P. C. Hohenberg and B. I. Halperin, Rev. Mod. Phys. {\bf 49}, 435 (1977).

\bibitem{Alexandrowicz-90}
Z. Alexandrowicz, Physica A {\bf 167}, 322 (1990).

\bibitem{vanHove-54}
L. van Hove, Phys. Rev. {\bf 93}, 1374 (1954).

\bibitem{Abe-69}
R. Abe and A. Hatano, Prog. Theor. Phys. {\bf 41}, 941 (1969).

\bibitem{Yahata-Suzuki-69}
H. Yahata and M. Suzuki, J. Phys. Soc. Jpn. {\bf 27}, 1421 (1969). 

\bibitem{Yahata-71}
H. Yahata, J. Phys. Soc. Jpn. {\bf 30}, 657 (1971).

\bibitem{Racz-Collins-76}
Z. R\'acz and M. F. Collins,  Phys. Rev. B {\bf 13}, 3074 (1976).

\bibitem{Rogiers-Indekeu-90} 
J. Rogiers and J. O. Indekeu,  Phys. Rev. B {\bf 41}, 6998 (1990).

\bibitem{Wang-93}
J. Wang, Phys. Rev. B {\bf 47}, 869 (1993).

\bibitem{Dammann-Reger-93}
B. Dammann and J. D. Reger, Europhys. Lett. {\bf 21}, 157 (1993);
Z. Phys. B {\bf 98}, 97 (1995). 

\bibitem{Adler-96}
J. Adler, in {\sl Annual Reviews of Computational Physics IV}, 
edited by D. Stauffer, p.~241 (World Scientific, Singapore, 1996).

\bibitem{Halperin-72}
B. I. Halperin, P. C. Hohenberg, and S. Ma, Phys. Rev. Lett. {\bf 29},
1548 (1972). 

\bibitem{DeDominicis-75}
C. De Dominicis, E. Br\'ezin, and J. Zin-Justin, Phys. Rev. B {\bf 12},
4945 (1975).

\bibitem{Bausch-81}
R. Bausch, V. Dohm, H. K. Janssen, and R. K. P. Zia, Phys. Rev. Lett
{\bf 47}, 1837 (1981). 

\bibitem{Achiam-80}
Y. Achiam, J. Phys. A {\bf 13}, 1355 (1980).

\bibitem{Mazenko-Valls-81}
G. F. Mazenko and O. T. Valls, Phys. Rev. B {\bf 24}, 1419 (1981);
{\it ibid.} {\bf 31}, 1565 (1985).

\bibitem{Takano-Suzuki-82}
H. Takano and M. Suzuki, Prog. Theor. Phys. {\bf 67}, 1332 (1982).

\bibitem{Haake-Lewenstein-Wilkens-84}
F. Haake, M. Lewenstein, and M. Wilkens, Z. Phys. B {\bf 54}, 333 (1984).

\bibitem{Tobochnik-81}
J. Tobochnik, S. Sarker, and R. Cordery, 
Phys. Rev. Lett. {\bf 46}, 1417 (1981). 

\bibitem{Kalle-84}
C. Kalle, J. Phys. A {\bf 17}, L801 (1984).

\bibitem{Williams-85}
J. K. Williams, J. Phys. A {\bf 18}, 49 (1985). 

\bibitem{Lacasse-93}
M.-D. Lacasse, J. Vi\~nals, and M. Grant, Phys. Rev. B {\bf 47}, 5646 (1993). 

\bibitem{MCRG}
S.-K. Ma, Phys. Rev. Lett. {\bf 37}, 461 (1976); 
R. H. Swendsen, Phys. Rev. Lett. {\bf 42}, 859 (1979);
R. H. Swendsen, in {\sl Real Space Renormalization}, edited by 
T. W. Burkhardt and
J. M. J. van Leeuwen (Springer, Berlin, 1982).

\bibitem{Stoll-Binder-Schneider-73}
E. Stoll, K. Binder, and T. Schneider, Phys. Rev. B {\bf 8}, 3266 (1973).

\bibitem{Tang-Wansleben-87}
S. Tang and D. P. Landau, Phys. Rev. B {\bf 36}, 567 (1987);
S. Wansleben and D. P. Landau, Phys. Rev. B {\bf 43}, 6006 (1991).

\bibitem{Landau-94}
D. P. Landau, Physica A {\bf 205}, 41 (1994). 

\bibitem{Pearson-85}
R. B. Pearson, J. L. Richardson, and D. Toussaint, 
Phys. Rev. B {\bf 31}, 4472 (1985).

\bibitem{Heuer-92}
H.-O. Heuer, J. Phys. A {\bf 25}, L567 (1992);
in {\sl Annual Reviews of Computational Physics IV}, edited by
D. Stauffer, p.~267 (World Scientific, Singapore, 1996).

\bibitem{Kikuchi-Okabe-86}
M. Kikuchi and Y. Okabe, J. Phys. Soc. Jpn, {\bf 55}, 1359 (1986).

\bibitem{Bonfim-87}
O. F. de Alcantara Bonfim, Europhys. Lett. {\bf 4}, 373 (1987). 

\bibitem{Mori-Tsuda-88}
M. Mori and Y. Tsuda, Phys. Rev. B {\bf 37}, 5444 (1988). 

\bibitem{Stauffer-92}
D. Stauffer, Physica A {\bf 184}, 201 (1992); 
{\it ibid.} {\bf 186}, 197 (1992);
Int. J. Mod. Phys. C {\bf 3}, 1059 (1992);
G. A. Kohring and D. Stauffer, Int. J. Mod. Phys. C {\bf 3}, 1165 (1992).

\bibitem{Muenkel-93}
C. M\"unkel, D. W. Heermann, J. Adler, M. Gofman, and D. Stauffer,
Physica A {\bf 193}, 540 (1993);
C. M\"unkel, Int. J. Mod. Phys. C {\bf 4}, 1137 (1993);
A. Linke, D. W. Heermann, P. Altevogt, and M. Siegert,
Physica A {\bf 222}, 205 (1995).

\bibitem{Ito-93}
N. Ito, Physica A {\bf 196}, 591 (1993); 
{\it ibid.} {\bf 192}, 604 (1993).

\bibitem{Stauffer-93}
D. Stauffer, Phys. A {\bf 26}, L599 (1993). 

\bibitem{Matz-94}
R. Matz, D. L. Hunter, and N. Jan, J. Stat. Phys. {\bf 74}, 903 (1994).

\bibitem{Grassberger-95}
P. Grassberger, Physica A {\bf 214}, 547 (1995). 

\bibitem{Gropengiesser-95}
U. Gropengiesser, Physica A {\bf 215}, 308 (1995).

\bibitem{Wang-Suzuki-Hu-95}
F. Wang, N. Hatano, and M. Suzuki, J. Phys. A {\bf 28} 4543 (1995).
F. Wang and M. Suzuki, Physica A {\bf 220}, 534 (1995).
F.-G. Wang and C.-K. Hu, Phys. Rev. E {\bf 56}, 2310 (1997).

\bibitem{Kikuchi-Ito-Okabe-94}
M. Kikuchi, N. Ito, and Y. Okabe, in {\sl Computer Simulation Studies
in Condensed-Matter Physics VII}, edited by D. P. Landau, K. K. Mon, and
H. B. Sch\"uttler, p.~44 (Springer, Berlin, 1994). 
 
\bibitem{Li-Schuelke-Zhang-95}
Z. B. Li, L. Sch\"ulke, and B. Zheng, Phys. Rev. Lett. {\bf 74}, 3396 (1995). 

\bibitem{Soares-97}
M. S. Soares, J. K. L. da Silva, and F. C. S\'a Barreto, 
Phys. Rev. B {\bf 55}, 1021 (1997). 

\bibitem{Pandit-Forgacs-Rujan-81}
R. Pandit, G. Forgacs, and P. Rujan, Phys. Rev. B {\bf 24}, 1576 (1981).

\bibitem{Nightingale-Bloete-96}
M. P. Nightingale and H. W. J. Bl\"ote, Phys. Rev. Lett. {\bf 76}, 4548 (1996).

\bibitem{Glauber-63}
R. J. Glauber, J. Math. Phys. {\bf 4}, 294 (1963).

\bibitem{Onsager-44}
L. Onsager, Phys. Rev. {\bf 65}, 117 (1944).

\bibitem{MullerKrumbhaar-73}
H. M\"uller-Krumbhaar and K. Binder, J. Stat. Phys. {\bf 8}, 1 (1973).

\bibitem{Swendsen-Wang-87}
R. H. Swendsen and J.-S. Wang, Phys. Rev. Lett. {\bf 58}, 86 (1987);
J.-S. Wang and R. H. Swendsen, Physica A {\bf 167}, 565 (1990).

\bibitem{Gan-Wang}
C. K. Gan and J.-S. Wang, J. Phys. A {\bf 29}, L177 (1996);
Phys. Rev. E {\bf 55}, 107 (1997); 
J. Chem. Phys. (to appear).

\bibitem{Privman-90} V. Privman, ed., {\sl Finite Size Scaling and the
Numerical Simulation of Statistical Systems} (Singapore, Word Scientific, 1990).

\bibitem{Suzuki-77} M. Suzuki, Prog. Theor. Phys. {\bf 58}, 1142 (1977).

\bibitem{Domany-84} E. Domany, Phys. Rev. Lett. {\bf 52}, 871 (1984).

\bibitem{Baker-Graves-Morris-81}
G. A. Baker and P. Graves-Morris, {\sl Pad\'e Approximants, Encyclopedia
of Mathematics and its Applications}, Vol.~13, and Vol.~14 (Reading, Mass.
Addison-Wesley, 1981).  

\bibitem{Guttmann-89}
A. J. Guttmann, in {\sl Phase Transitions and Critical Phenomena},
Vol.~13, edited by C. Domb and J. L. Lebowitz (Academic, New
York, 1989).

\bibitem{Dickman-91}
R. Dickman, J.-S. Wang, I. Jensen, J. Chem. Phys. {\bf 94}, 8252 (1991);
I. Jensen and R. Dickman, J. Stat. Phys. {\bf 71}, 89 (1993).

\bibitem{Stauffer-95}
D. Stauffer, Physica A {\bf 215}, 305 (1995).

\end{thebibliography}

\clearpage
\begin{table}
\caption{CPU time and memory usage for the series expansion
of relaxation of magnetization, measured on an AlphaStation
250/266.}
\label{tab:complexity}
\begin{tabular}{rrrr}
$n$ & \quad CPU time (sec) & Memory (MB) \\
\hline\hline
6  &  0.13 & 0.03 \\
7  &  1.8 & 0.27 \\
8  &  25  & 3   \\
9  &  358 & 34  \\
10 &  23600 & 51 \\
11 &  939000 & 70 \\
12\tablenote{Actual computations are done on a 16-node Pentium Pro 200 cluster.} 
   & $1.6\times 10^7$ & 85 \\ 
\end{tabular}
\end{table}

\widetext
\begin{table}
\caption{Series-expansion coefficients ($n$-th derivative) 
for a single spin $\langle \sigma_0\rangle_t$ and nearest neighbor
spin correlation $\langle \sigma_0\sigma_1\rangle_t$.}
\label{tab:series}
\begin{tabular}{rrrr}
$n$ & \quad ${ d^n \langle \sigma_0\rangle_t \over dt^n} |_0$ &
${ d^n \langle \sigma_0\sigma_1\rangle_t \over dt^n}|_0$ \\ 
\hline\hline

0 & $1$  &  $1$ \\
 
1 & $-1 + (2\sqrt{2})/3$ &           $ -2 + (4\sqrt{2})/3 $ \\
 
2 & $13/9 - \sqrt{2}$ &              $ (56 - 39\sqrt{2})/9 $ \\
 
3 & $(15 - 11\sqrt{2})/27$ &         $ 2(-249 + 175\sqrt{2})/27 $ \\
 
4 & $-53/3 + 25/\sqrt{2} $ &         $ (1988 - 1399\sqrt{2})/54 $ \\
 
5 & $(41175 - 29111\sqrt{2})/486$ &  $ (30834 - 21919\sqrt{2})/486 $ \\
 
6 & $(-66133 + 46680\sqrt{2})/1458$ &
                                    $ 2(-142869 + 101087\sqrt{2})/243 $ \\
 
7 & $(-125718825 + 88903747\sqrt{2})/34992 $ &
                               $ 5(18191091 - 12867401\sqrt{2})/17496 $ \\
 
8 & $17(92513582 - 65418301\sqrt{2})/34992 $ &
                              $ (2190719830 - 1548846809\sqrt{2})/69984 $ \\
 
9 & $(-429437553903 + 303660675715\sqrt{2})/1259712 $ &
                        $ (-289028693217 + 204371192813\sqrt{2})/314928 $ \\
 
10 & $ (4931635327666 - 3487215692619\sqrt{2})/3779136 $ &
                    $ (43146864055759 - 30509318092215\sqrt{2})/3779136 $ \\
 
11 & $ (1821425391381531 - 1287938652305897\sqrt{2})/181398528 $ &
                $ (-957792089655213 + 677259915390707\sqrt{2})/10077696 $ \\
 
12 & $7(\!-\!10761633667757321\!\!+\!\!7609621330268025\sqrt{2})/272097792$ &
  $ (425962164223774298\!\! -\!\! 301200006005168631\sqrt{2})/1088391168 $ \\
\end{tabular}
\end{table}

\narrowtext

\begin{figure}
\caption{Pad\'e estimates of the dynamical critical exponent $z$
using $G_1(t=\infty)$, plotted as a function of $\Delta$, the transformation
parameter.  On this scale, the Pad\'e approximant of order $[N,D]$
is indistinguishable from $[D,N]$.}
\label{fig:Gb}
\end{figure}

\begin{figure}
\caption{Effective exponent $z_{\text{eff}}(t)$ plotted
against inverse time
$1/t$.   The circles are Monte Carlo estimates based on Eq.~(\ref{ratio});
the continuous curve is obtained from the $[6,6]$ Pad\'e approximant of $G_1$ in
variable $u$, transformed back to $F$ through Eq.~(\ref{FGrelation}).
}
\label{fig:zeff}
\end{figure}
\clearpage
 
\input epsf
\vfill
\epsfxsize\hsize\epsfbox{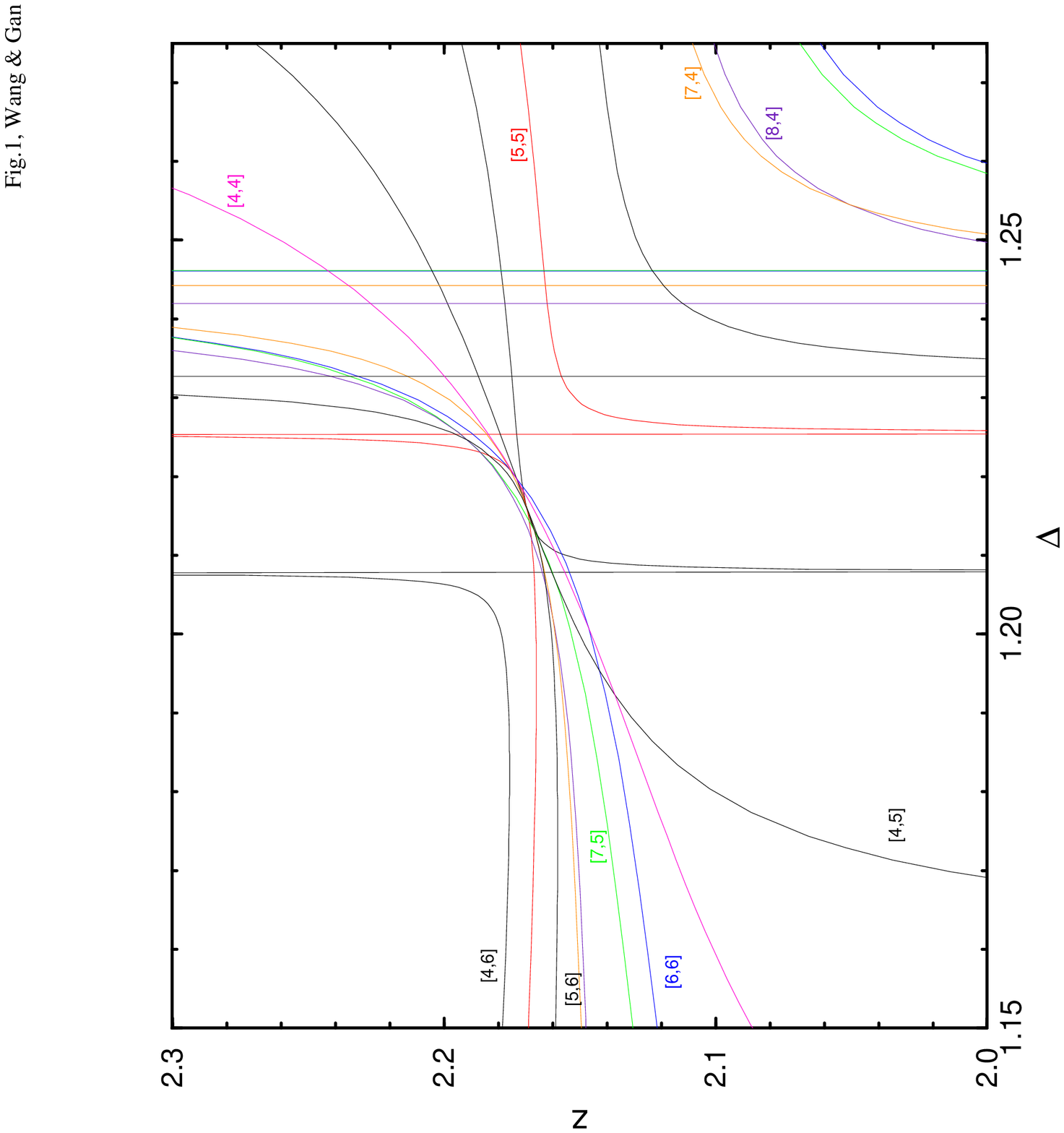}
\vfill
\epsfxsize\hsize\epsfbox{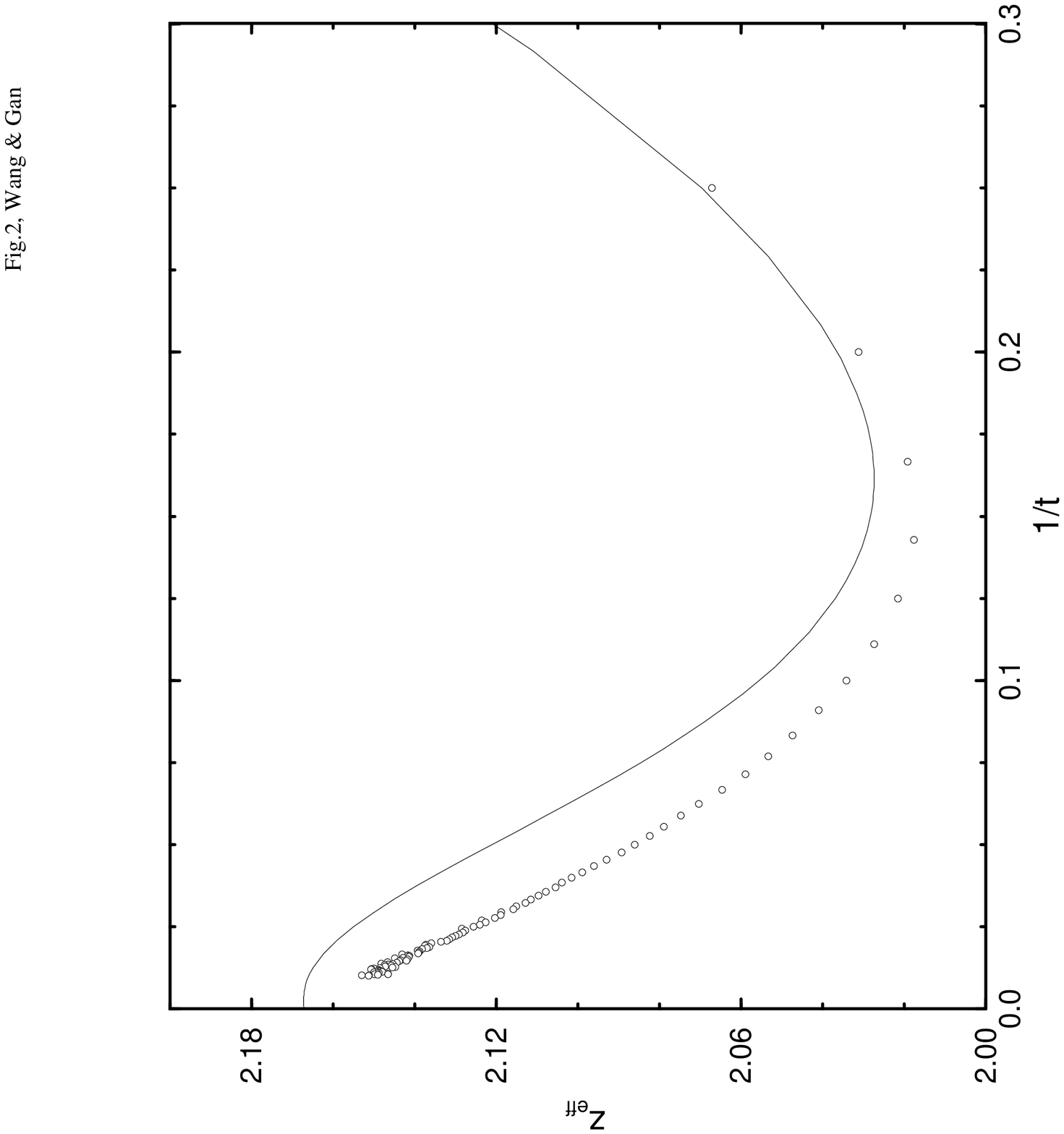}
\vfill

\end{document}